\documentclass[english,a4paper]{article}
\usepackage[T1]{fontenc}
\usepackage[cp1250]{inputenc}
\usepackage{amssymb}
\usepackage{hyperref}
\usepackage{amsfonts}
\usepackage{graphicx}

\begin{document}

\title{Simultaneous time-optimal control of the inversion of two spin 1/2 particles}

\author{E. Ass\'emat, M. Lapert, Y. Zhang, M. Braun, S. J. Glaser\footnote{Department of Chemistry, Technische Universit\"at M\"unchen, Lichtenbergstrasse 4, D-85747 Garching, Germany} and D. Sugny\footnote{Laboratoire Interdisciplinaire Carnot de Bourgogne (ICB), UMR 5209 CNRS-Universit\'e de Bourgogne, 9 Av. A. Savary, BP 47 870, F-21078 DIJON Cedex, FRANCE, dominique.sugny@u-bourgogne.fr}}

\maketitle

\begin{abstract}
We analyze the simultaneous time-optimal control of two-spin
systems. The two non coupled spins which differ in the value of
their chemical offsets are controlled by the same magnetic fields.
Using an appropriate rotating frame, we restrict the study to the
case of opposite shifts. We then show that the optimal solution of
the inversion problem in a rotating frame is composed of a pulse sequence of maximum intensity
and is similar to the optimal solution for inverting only one
spin by using a non-resonant control field in the laboratory frame. An example is implemented experimentally using techniques of
Nuclear Magnetic Resonance.
\end{abstract}
%\pacs{32.80.Qk,37.10.Vz,78.20.Bh} \maketitle
\section{Introduction}
Since its discovery in 1945 by Purcell, Torrey and Pound, Nuclear
Magnetic Resonance (NMR) has become a powerful physical tool to
study molecules and matter in a variety of domains extending from
biology and chemistry to solid physics and quantum mechanics
\cite{spin}. NMR involves the manipulation of nuclear spins
via its interaction with a magnetic field, and is therefore a
domain where techniques of quantum control can be applied (see
\cite{vander} and references therein). Such an approach has many
potential applications ranging from the improvement of the
resolution and sensitivity of NMR spectroscopy experiments
\cite{fruech} to quantum computing \cite{chuang}. The control technology developed over the past fifty years allows the use of sophisticated control fields for spectroscopy and also permits the implementation of complex quantum algorithms \cite{nielsen}.

In this context, some challenging control problems are raised by
the experimental constraints of NMR experiments. Roughly speaking,
the measured signal is the magnetization of a sample which is
produced by a large number of spin systems. One usually assumes in
simple models that the static magnetic field is the same across the
sample, i.e. the field is perfectly homogeneous with respect to
the different spins. This is not always true in practice since for
technical reasons it is difficult to generate homogeneous fields.
Even in the situation where the magnetic field is uniform on a
macroscopic scale, the interaction between the different atoms (or
between a spin and its environment) induces a chemical shift on
the frequency transition of a given spin. This leads classically
to an unwanted rotation of each individual spin around a fixed
axis which is not taken into account in the simplest model of spin
1/2 particles. The shift being different for each spin, the
rotation is different for each spin. Note that this effect is
useful in NMR spectroscopy since it encodes in a sense some
informations about the structure of the molecules. The
consequences are negative from a control point of view since this
phenomenon decreases the efficiency of the control field. The
objective is therefore to find controls able to bring the
system towards a given target state in a sufficiently robust way
with respect to inhomogeneities of the transition frequency. This
problem has been solved numerically in different works
\cite{grapeino} leading to very efficient but complicated
solutions. In particular, no insight into the control mechanism is
gained from this approach and no optimality result has been
proven. Note that some related works have been done in the control
of molecular dynamics by laser fields \cite{turinici} by using
monotonically convergent algorithms \cite{mono}.

In this paper, we propose to revisit this problem by using
techniques of geometric optimal control theory
\cite{jurdjevic,bonnard}. Geometric optimal control is a vast
domain based on the application of the Pontryagin Maximum
Principle (PMP) where the idea is to use the methods of
differential geometry and Hamiltonian dynamics to solve the
optimal control problems \cite{jurdjevic,bonnard}. This geometric
framework leads to a global analysis of the control problem which
completes and guides the numerical computations. Some geometric
results on the optimal control of spin systems have been first
obtained by N. Khaneja and his co-workers \cite{khaneja}.
Recently, the time-optimal control of dissipative spin 1/2
particles has been solved theoretically \cite{sugny1} and
implemented experimentally \cite{lapertglaser}. In this work
%, due
%to the complexity of the experimental sample which can be composed
%of thousand spin systems,
%we restrict the study of the
we study the simultaneous control of %spin systems to only two spins with
two non-interacting spins with
different resonance frequencies. More precisely, we consider as an
%example the inversion control problem of the magnetization vector
%along the $z$- axis, which is defined by the longitudinal magnetic
%field.
example the problem to simultaneously invert the magnetization vectors
initially aligned
along the $z$- axis defined by the direction of the static magnetic
field.

Using an appropriate rotating frame, we show that we can
always consider the symmetric case where the two transition
frequencies are opposed. In this situation, the time-optimal
solution for inverting the two spins by the same transverse radiofrequency (rf) control fields
is a bang-bang pulse sequence in a frame rotating at the rf frequency. The remarkable point is that the co-rotating component of the applied
rf field is the same as the one used to invert only one spin with
one non-resonant control field in the laboratory frame \cite{boscain2}. We finally implement experimentally the optimal
solution by using techniques of NMR.

The paper is organized as follows. In Sec. \ref{sec2}, we recall
the tools to control one spin in minimum time with a transverse
magnetic field which is not in resonance with the frequency of the
spin. In Sec. \ref{sec3}, we establish that this control field is
also the optimal solution to simultaneously invert two spins. An
experimental illustration is given in Sec. \ref{sec4}. A summary
of the different results obtained is presented in Sec. \ref{sec5}.
\section{Time-optimal control of a spin 1/2 particle}\label{sec2}
We consider the control of a spin 1/2 particle whose dynamics is governed by the Bloch equation:
\begin{equation}
\left(\begin{array}{c}
\dot{M}_x \\
\dot{M}_y \\
\dot{M}_z \end{array}\right)
=\left(\begin{array}{c}
-\omega M_y \\
\omega M_x \\
0
\end{array}\right) +
\left(\begin{array}{c}
0 \\
-\omega_{x} M_z \\
\omega_{x} M_y
\end{array}\right)
\end{equation}
where $\vec{M}$ is the magnetization vector and $\omega$
the chemical shift offset. The dynamics is controlled through only
one magnetic field along the $x$- axis which satisfies the
constraint $\omega_x\leq \omega_{max}$. We introduce normalized
coordinates $\vec{x}=(x,y,z)=\vec{M}/M_0$ where
$\vec{M}_0=M_0\vec{e}_z$ is the thermal equilibrium point, a
normalized control field $u_x=2\pi\omega_x/\omega_{max}$ which
satisfies the constraint $u_x\leq 2\pi$ and a normalized time
$\tau=\omega_{max}t/(2\pi)$. Dividing the previous system by
$\omega_{max}M_0/(2\pi)$, we get that the evolution of the
normalized coordinates is given by the following equations:
\begin{equation}
\left(\begin{array}{c}
\dot{x} \\
\dot{y} \\
\dot{z} \end{array}\right)
=\left(\begin{array}{c}
-\Delta y \\
\Delta x \\
0
\end{array}\right) +
\left(\begin{array}{c}
0 \\
-u_x z \\
u_x y
\end{array}\right)
\end{equation}
where $\Delta$ is the normalized offset given by $\Delta=2\pi\omega/\omega_{max}$.

The complete description of the time-optimal control problem of a
spin 1/2 particle by a non-resonant magnetic field is done in Ref.
\cite{boscain2}. In this section, we give only a brief summary of
the results of this paper which will be used in our study. The
reader is referred to \cite{boscain2} for the different proofs of
these results. Note that when the spin is controlled by two
magnetic fields, respectively along the $x$ and $y$ directions, then the
system is equivalent to a two-level quantum system in the rotating
wave approximation \cite{boscain1}. This means that a unitary
transformation can be used to remove the drift term depending on
$\Delta$. In this case, the optimal control field is a
$\pi$-pulse.

The problem we consider belongs to a general class of optimal
control problems for which powerful mathematical tools have been
developed \cite{boscainbook}. They correspond to systems on a
two-dimensional manifold (here the Bloch sphere) controlled by
a single field. The evolution of the system is ruled by the
following set of differential equations:
\begin{equation}
\dot{\vec{x}}=F(\vec{x})+uG(\vec{x})
\end{equation}
where $\vec{x}$ is the two-dimensional state vector and $u$ the
control field which satisfies the constraint $u\leq u_0$ with,
here, $u_0=2\pi$. The time-optimal control problem is solved by
the application of the Pontryagin Maximum Principle (PMP) which is
formulated using the pseudo-Hamiltonian
\[
H=\vec{p}\cdot(F+uG)+p_{0},\]
where $\vec{p}$ is the adjoint state and $p_{0}$
a negative constant such that $\vec{p}$ and $p_0$ are not simultaneously equal to 0. The Pontryagin maximum principle states that
the optimal trajectories are solutions of the equations
\begin{eqnarray}
\begin{array}{lll}
\dot{\vec{x}}=\frac{\partial
H}{\partial\vec{p}}(\vec{x},\vec{p},v),\
\dot{\vec{p}}=-\frac{\partial H}{\partial\vec{x}}(\vec{x},\vec{p},v)\\
H(\vec{x},\vec{p},v)=\max_{|u|\leq u_0}H(\vec{x},\vec{p},u)\\
H(\vec{x},\vec{p},v)=0.\end{array}\label{eqpont}
\end{eqnarray}
Introducing the switching function $\Phi$ given by
\[
\Phi(t)=\vec{p}\cdot G,
\]
one deduces using the second equation of (\ref{eqpont}) that the
optimal synthesis is composed of concatenation of arcs
$\gamma_{+}$, $\gamma_{-}$ and $\gamma_{S}$. $\gamma_{+}$ and
$\gamma_{-}$ are regular arcs corresponding respectively to
$\textrm{sign}[\Phi(t)]=\pm 1$ or to the control fields $u=\pm
u_0$. A switching from $u_0$ to $-u_0$ or from $-u_0$ to $u_0$
occurs at $t=t_0$ when the function $\Phi$ takes the value zero
and when this zero is isolated. Singular arcs $\gamma_{s}$ are
characterized by the fact that $\Phi$ vanishes on an interval
$[t_{0},t_{1}]$. In this case, differentiating two times $\Phi$
with respect to time and imposing that the derivatives are zero,
one obtains that the singular arcs are located in the set \[
S=\{\vec{x};\ \Delta_{S}(x)=\det(G,[G,F])(\vec{x})=0\}.\] We
recall that the commutator $[F,G]$ of two vector fields $F$ and
$G$ is defined by:
\[[F,G]=\nabla F\cdot G-F\cdot \nabla G
\]
where $\nabla$ is the gradient of a function. The singular control
field $u_{s}$ can be calculated as a feedback control, i.e. as a
function of the coordinates by imposing that the second derivative
of $\Phi$ with respect to time is equal to 0:
\[
[G,[G,F]]+u_{s}[F,[G,F]]=0.\] The optimal solution can follow the
singular lines if the control field is admissible, i.e. if
$|u_{s}(\vec{x})|\leq u_0$.

Since the two-dimensional manifold of our control problem is the Bloch sphere, the adapted coordinates are the spherical ones:
\begin{equation}
\left\{\begin{array}{rcl}
x & = & r\,\sin\theta \, \cos\phi  \\
y & = & r\,\sin\theta \, \sin\phi \\
z & = & r\,\cos\theta
\end{array}\right. ,
\end{equation}
which leads to the following system:
\begin{equation}
\left(\begin{array}{c}
\dot{r} \\
\dot{\theta} \\
\dot{\phi}
\end{array}\right) = \left(\begin{array}{c}
0 \\
0 \\
\Delta
\end{array}\right) + u\left(\begin{array}{c}
0 \\
-\sin\,\phi \\
-\cos\,\phi\,\cot\,\theta.
\end{array}\right)
\end{equation}
The pseudo-Hamiltonian $H$ has the form:
\begin{equation}
H = \Delta\,p_{\phi} - u(\sin\,\phi \, p_{\theta} + \cos\,\phi \, \cot\,\theta\, p_{\phi}).
\end{equation}
and the switching function is given by $\Phi = \sin \phi p_{\theta} + \cos\, \phi\,\cot\,\theta\,p_{\phi}$. Since
\[[G,F]= \left(\begin{array}{c}
0 \\
-\Delta\,\cos\,\phi \\
\Delta\,\sin\,\phi\,\cot\,\theta
\end{array}\right)
\]
one deduces that $S$ is the set
\[S=\left\{\vec{x}|\sin^2\phi\cot\theta = -\cos^2\phi\cot\theta\right\} = \left\{\vec{x}|\theta = \pi/2\right\},\]
i.e. the singular locus is the equator of the Bloch sphere. The
time-optimal control problem is solved in \cite{boscain2}. It has
been shown that the optimal solution reaching the south pole from
the north pole is the succession of different bang pulses, i.e. of pulses of maximum intensity $2\pi$. The number of
bangs is at most equal to 2 if $\Delta<2\pi$ and can be larger if
$\Delta>2\pi$. The singular extremals play no role for this spin
inversion. An example of an optimal pulse sequence and the corresponding trajectory  is displayed in Figs. \ref{fig1} and \ref{fig2}. The optimal
trajectory is not smooth at switching points where the value of
the control field changes. The switching times can analytically be
determined using the material of Ref. \cite{boscain2}. In the case
of Fig. \ref{fig1}, the optimal solution is a type-2-trajectory
described by Proposition 5 of \cite{boscain2}. These times can
also be computed numerically by solving a shooting equation. More
precisely, this means that one has to determine the initial
adjoint state $\vec{p}(0)=(p_\theta(0),p_\phi(0))$ such that the
corresponding Hamiltonian trajectory $(\vec{x},\vec{p})$ with initial conditions
$(\vec{x}(0),\vec{p}(0))$ goes to the target $\vec{x}_f$ at time
$t_f$. This condition can be expressed as the determination of the
roots of the equation $\vec{x}(t_f)[\vec{p}(0)]-\vec{x}_{f}$
which can be solved if one has a sufficiently good approximation
of $\vec{p}(0)$ by a standard Newton-type algorithm. Note that the
control field is determined along the trajectory by computing the
switching function $\Phi$.

At this point, we can extend the previous discussion as a first step towards the simultaneous
inversion of two spins. We analyze the dynamics in a rotating frame by using the rotating wave approximation (RWA) where the offsets of the two spins are symmetric and given by
$\pm \Delta$. The rf field is assumed to be at the rotating frame frequency. As a consequence of the
symmetries of the problem, one sees that if $u(t)$, the co-rotating component of the applied rf field, steers the spin
with offset $\Delta$ from the north pole to the south pole then
the same field will also invert the other spin. The trajectories
of the two spins in the $y$- and $z$- directions will be the same,
while it will be opposite along the $x$- axis. Note that this
solution is not the unique solution and a family of solutions
satisfying the same requirement can be determined. Consider the
set of control fields defined by
\begin{equation}
\left\{\begin{array}{rcl}
u_x=u(t)\cos\alpha\\
u_y=u(t)\sin\alpha\\
   \end{array}\right.
\end{equation}
where $\alpha\in [0,2\pi]$. If we consider the following rotation
$R(\alpha)$ of angle $\alpha$ along the $z$- axis:
\begin{equation}
\left(\begin{array}{c}
X \\
Y \\
Z         \\
\end{array}\right)
= \left(\begin{array}{ccc}
\cos\alpha & \sin\alpha & 0 \\
-\sin\alpha & \cos\alpha & 0 \\
0 & 0 & 1 \\
\end{array}\right)
\left(\begin{array}{c}
x \\
y \\
z         \\
\end{array}\right)
\end{equation}
then the new system in the coordinates $(X,Y,Z)$ is controlled by
a single field $u(t)$ along the $X$- direction. It is also straightforward to see that this solution is the
optimal one for the inversion control of two symmetric spins by one control field.

The question that we ask now is whether this simple solution is the
optimal solution of the simultaneous inversion of two spins when two control fields are considered.
\section{Simultaneous control of the inversion of two spin 1/2
particles}\label{sec3}
\subsection{The model system}
We consider two different spin 1/2 particles with the offsets
$\omega_a$ and $\omega_b$. Using the same
normalization as in Sec. \ref{sec2} and the RWA, one arrives to the following
equations:
\begin{equation}
\left(\begin{array}{c}
\dot{x}_a \\
\dot{y}_a \\
\dot{z}_a \\
\dot{x}_b \\
\dot{y}_b \\
\dot{z}_b
\end{array}\right) = \left(\begin{array}{c}
-\Delta_a y_a \\
\Delta_a x_a \\
0         \\
-\Delta_b y_b \\
\Delta_b x_b \\
0
\end{array}\right) + u_x\left(\begin{array}{c}
0 \\
- z_a \\
y_a  \\
0     \\
- z_b \\
y_b
\end{array}\right) + u_y\left(\begin{array}{c}
z_a \\
0 \\
- x_a \\
z_b \\
0 \\
- x_b
\end{array}\right)
\end{equation}
where the coordinates $(x_a,y_a,z_a)$ and $(x_b,y_b,z_b)$ are
respectively associated to the first and second spins $a$ and $b$. The parameters $\Delta_a$ and $\Delta_b$ are the offsets of the spins $a$ and $b$ with respect to the frequency of the rotating frame. The rf field is here also at the rotating frame frequency. We assume
that the two spins have the same equilibrium point $M_0$. As mentioned below, two
magnetic fields along the $x$- and $y$- directions are taken into
account in this problem. They satisfy the constraints
$\sqrt{u_x^2+u_y^2}\leq 2\pi$. Using a rotating frame that rotates
at frequency $(\Delta_a+\Delta_b)/2$, it is straightforward to
transform this system into a symmetric one where the frequencies
of the two spins are opposite. This is the case analyzed below.

We introduce the spherical coordinates for the two spins and we
get:
\begin{equation}
\left(\begin{array}{c}
\dot{r_a} \\
\dot{\theta_a} \\
\dot{\phi_a} \\
\dot{r_b} \\
\dot{\theta_b} \\
\dot{\phi_b} \\
\end{array}\right) = \left(\begin{array}{c}
0 \\
0 \\
\Delta \\
0 \\
0 \\
-\Delta
\end{array}\right) + u_xG_x + u_yG_y
\end{equation}
where
\[
G_x=\left(\begin{array}{c}
0 \\
-\sin\phi_a \\
-\cot\theta_a \, \cos\phi_a \\
0 \\
-\sin\phi_b \\
-\cot\theta_b \, \cos\phi_b
\end{array}\right),~G_y=\left(\begin{array}{c}
0 \\
\cos\,\phi_a \\
-\cot\theta_a \, \sin\phi_a \\
0 \\
\cos\,\phi_b \\
-\cot\theta_b \, \sin\phi_b
\end{array}\right).\]
Since the radial coordinates $(r_a,r_b,p_{r_a},p_{r_b})$ play a
trivial role in this problem, we omit them in the following
equations.

We apply the PMP to this system in the time-optimal case and we
obtain the following pseudo-Hamiltonian
\begin{equation}
H=\Delta(p_{\phi_a}-p_{\phi_b})+\vec{p}\cdot
(u_xG_x+u_yG_y)
\end{equation}
where $\vec{p}$ is the adjoint vector of coordinates
$(p_{\theta_a},p_{\phi_a},p_{\theta_b},p_{\phi_b})$. In the normal
case, the optimization condition leads to the following optimal
controls:
\begin{equation}
u_x=\frac{\vec{p}\cdot G_x}{\sqrt{(\vec{p}\cdot
G_x)^2+(\vec{p}\cdot G_y)^2}},~u_y=\frac{\vec{p}\cdot
G_y}{\sqrt{(\vec{p}\cdot G_x)^2+(\vec{p}\cdot G_y)^2}},
\end{equation}
where $\vec{p}\cdot G_x$ and $\vec{p}\cdot G_y$ are not
simultaneously equal to 0. The singular case occurs when
$\vec{p}\cdot G_x=\vec{p}\cdot G_y=0$, which defines the
switching surface $\Sigma$. In the two-control problems, singular
trajectories are the trajectories which lie on $\Sigma$. We assume
in this paper that these controls do not play any role in our
problem. This is expected since singular extremals are not
generically optimal for a two-control problem \cite{chitour}.

We get the normal Hamiltonian $H_n$ by replacing the control
fields by their expressions:
\begin{equation}
H_n=\Delta(p_{\phi_a}-p_{\phi_b})+\sqrt{(\vec{p}\cdot
G_x)^2+(\vec{p}\cdot G_y)^2}.
\end{equation}
The normal extremals will be given by the Hamiltonian trajectories
of $H_n$. The next step of our study will consists in the analysis
of this Hamiltonian flow.

For that purpose, we introduce the following canonical transformation on the $\phi$- coordinates:
\begin{equation}
\left\{\begin{array}{rcl}
\phi_+ & = & \phi_a + \phi_b \\
\phi_- & = & \phi_a - \phi_b \\
\end{array}\right.
\end{equation}
which is defined through the generating function:
\[
F_2=\frac{1}{2}p_{\phi_a}(\phi_++\phi_-)+\frac{1}{2}p_{\phi_b}(\phi_+-\phi_-)
\]
with the transformation:
\[
p_{\phi_+}=\frac{\partial F_2}{\partial \phi_+};~
p_{\phi_-}=\frac{\partial F_2}{\partial \phi_-};~
\phi_a=\frac{\partial F_2}{\partial p_{\phi_a}};~
\phi_b=\frac{\partial F_2}{\partial p_{\phi_b}}.
\]
This leads to
\begin{equation}
\left\{\begin{array}{rcl}
p_{\phi_a} & = & p_{\phi_+} + p_{\phi_-} \\
p_{\phi_b} & = & p_{\phi_+} - p_{\phi_-} \\
\end{array}\right. .
\end{equation}
The Hamiltonian $H_n$ expressed in the new set of coordinates does
not depend of $\phi_+$, so $p_{\phi_+}$ is a constant of the
motion. Since at the initial time in the north pole,
$p_{\phi_a}(0)=p_{\phi_b}(0)=0$, one deduces that $p_{\phi_+}=0$.
One finally arrives at
\begin{eqnarray*}
& & H_n=2\Delta p_{\phi_-}+[p_{\theta_a}^2+p_{\theta_b}^2+p_{\phi_-}^2(\cot^2\theta_a+\cot^2\theta_b)\\
& & +2\cos\phi_-(p_{\theta_a}p_{\theta_b}-p_{\phi_-}^2\cot\theta_a\cot\theta_b)\\
& & -2p_{\phi_-}\sin\phi_-(p_{\theta_a}\cot\theta_b-p_{\theta_b}\cot\theta_a)]^{1/2}.
\end{eqnarray*}
Care has to be taken with the use of these coordinates on the poles
of the sphere. On a pole, we have $\cot\theta\to \pm \infty$ and
$p_\phi=0$ but the product $p_\phi\cot\theta$ remains finite. In
this paper, spherical coordinates are only used to describe the
geometric properties of the extremals and to highlight their
symmetries. All the numerical computations are done in cartesian
coordinates.

Note also the symmetric role played by $\theta_a$ and $\theta_b$ in
the Hamiltonian $H_n$. This symmetry will be used in the proof
below.
\subsection{The optimal control problem}
We first analyze the characteristics of the extremal trajectories which are solutions of the control problem. In particular, if the inversion is realized by an extremal trajectory then the following relations are satisfied:
\[
\forall t\in[0,t_f],
p_{\theta_a}(t)=p_{\theta_b}(t)~\textrm{and}~\theta_a(t)=\theta_b(t),
\]
where $t_f$ is the control duration.

To show this property, we assume that the
south pole is reached by the extremal. In this case, the final point satisfies
by definition:
\[
\theta_a(t_f)=\theta_b(t_f),~\dot{\theta}_a(t_f)=\dot{\theta}_b(t_f),
p_{\phi_-}=0.
\]
Using the Hamiltonian $H_n$, we obtain:
\begin{equation}\label{eqder}
\left\{\begin{array}{rcl}
\dot{\theta_a}=\frac{\partial H_n}{\partial
p_{\theta_a}}=(p_{\theta_a}+\cos\phi_-p_{\theta_b}-p_{\phi_-}\sin\phi_-\cot\theta_b)/\sqrt{Q}\\
\dot{\theta_b}=\frac{\partial H_n}{\partial
p_{\theta_b}}=(p_{\theta_b}+\cos\phi_-p_{\theta_a}-p_{\phi_-}\sin\phi_-\cot\theta_a)/\sqrt{Q}\\
   \end{array}\right. ,
\end{equation}
where
\begin{eqnarray*}
& &
Q=p_{\theta_a}^2+p_{\theta_b}^2+p_{\phi_-}^2(\cot^2\theta_a+\cot^2\theta_b)\\
& &
+2\cos\phi_-(p_{\theta_a}p_{\theta_b}-p_{\phi_-}^2\cot\theta_a\cot\theta_b)\\
& &
-2p_{\phi_-}\sin\phi_-(p_{\theta_a}\cot\theta_b-p_{\theta_b}\cot\theta_a).
\end{eqnarray*}
We consider that $Q(0)\neq 0$ which is always possible by a
judicious choice of the initial adjoint vector $\vec{p}(0)$. This
implies that $H_n>0$ since $p_{\phi_-}=0$ at a pole. Using the
fact that $H_n$ is a constant of motion, one deduces at the final
point that $Q(t_f)\neq 0$.  From Eqs. (\ref{eqder}), one finally
arrives at
\[
(p_{\theta_a}(t_f)-p_{\theta_b}(t_f))(1-\cos\phi_-(t_f))=0.
\]
If $\cos\phi_-(t_f)=1$ then $Q=0$, which is not possible from our
hypothesis. We have therefore
$p_{\theta_a}(t_f)=p_{\theta_b}(t_f)$ and using the Hamiltonian
equations, we then obtain that $\theta_a(t)=\theta_b(t)$ and
$p_{\theta_a}(t)=p_{\theta_b}(t)$ for any time $t$ since at time
$t=t_f$ we have $\theta_a(t_f)=\theta_b(t_f)$ and
$p_{\theta_a}(t_f)=p_{\theta_b}(t_f)$.

%To ensure that $z_1(t)=z_2(t)$, the initial adjoint states have to
%satisfy $|p_{x_1}(0)|=p_{x_2}(0)|$ and $|p_{y_1}(0)|=p_{y_2}(0)|$.

%\subsection{The symmetric case}
%In this section, we consider the symmetric situation for which
%$\Delta_1=-\Delta_2$. From an experimental point of view, it is
%always possible to be in this case by choosing adequately the
%frequency of the transverse magnetic field.

We also get that $\dot{\phi}_+=0$, i.e. $\phi_+=\phi_{+0}$. In the
new coordinates $\vec{X}_a$ and $\vec{X}_b$ such that
$\vec{X}_a=R(\phi_{+0}/2)\vec{x}_a$ and
$\vec{X}_b=R(\phi_{+0}/2)\vec{x}_b$, the sum of the new azimuthal
angles is zero and we obtain the following symmetry on the
trajectory:
\begin{equation}
\left\{\begin{array}{rcl}
       X_a(t) & = & X_b(t) \\
       Y_a(t) & = & -Y_b(t) \\
       Z_a(t) & = & Z_b(t)
   \end{array}\right.
\end{equation}
for any $t\in [0,t_f]$. In these coordinates, the two control
fields are given by
\begin{equation}
\left\{\begin{array}{rcl}
       u_X(t) & = & u_x(t)\cos(\frac{\phi_0}{2})+u_y(t)\sin(\frac{\phi_0}{2}) \\
       u_Y(t) & = & -u_x(t)\sin(\frac{\phi_0}{2})+u_y(t)\cos(\frac{\phi_0}{2}) \\
       \end{array}\right.
\end{equation}
From this symmetry, we have $u_X(t)=0$ and thus
$u_x(t)\cos(\phi_{0+}/2)+u_y(t)\sin(\phi_{0+}/2)=0$. Since
$u_x(t)^2+u_y(t)^2\leq 4\pi ^2$, this leads to
\begin{equation}
\left\{\begin{array}{rcl}
       u_x(t)=u_0(t)\cos(\phi_{0+}/2) \\
       u_y(t)=-u_0(t)\sin(\phi_{0+}/2)
   \end{array}\right.
\end{equation}
where $u_0(t)$ is a bang-bang pulse of amplitude $2\pi$. We
therefore recover the case of Sec. \ref{sec2} of the optimal
control of one spin system.
\begin{figure}[htbp]
\begin{center}
\includegraphics[height=4.5in]{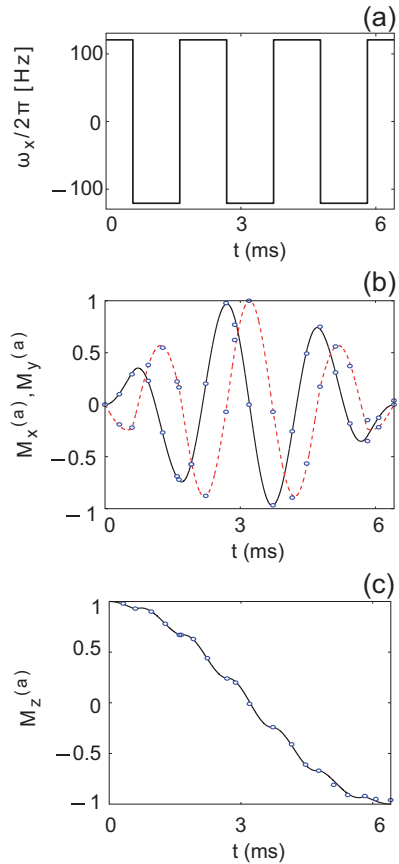}
\end{center}
\caption{(Color Online) Panel (a): Plot of the optimal control field for
the simultaneous inversion of two spins with offsets $\omega/2 \pi=\pm 483$ Hz and
$\omega_{max}/ 2 \pi=120.75$ Hz. Panel (b):  Plots of the corresponding optimal trajectories for $M_x^{a}(t)$ (solid black line) and $M_y^{a}(t)$ (red dashed line). The experimentally measured trajectories are presented by open circles.
Panel (c): Plot of the optimal trajectory for $M_z^{a}(t)$, together with the experimental representations (open circles). \label{fig1}}
\end{figure}

\section{Experimental illustration}\label{sec4}
\begin{figure}
\begin{center}
\includegraphics[height=2in]{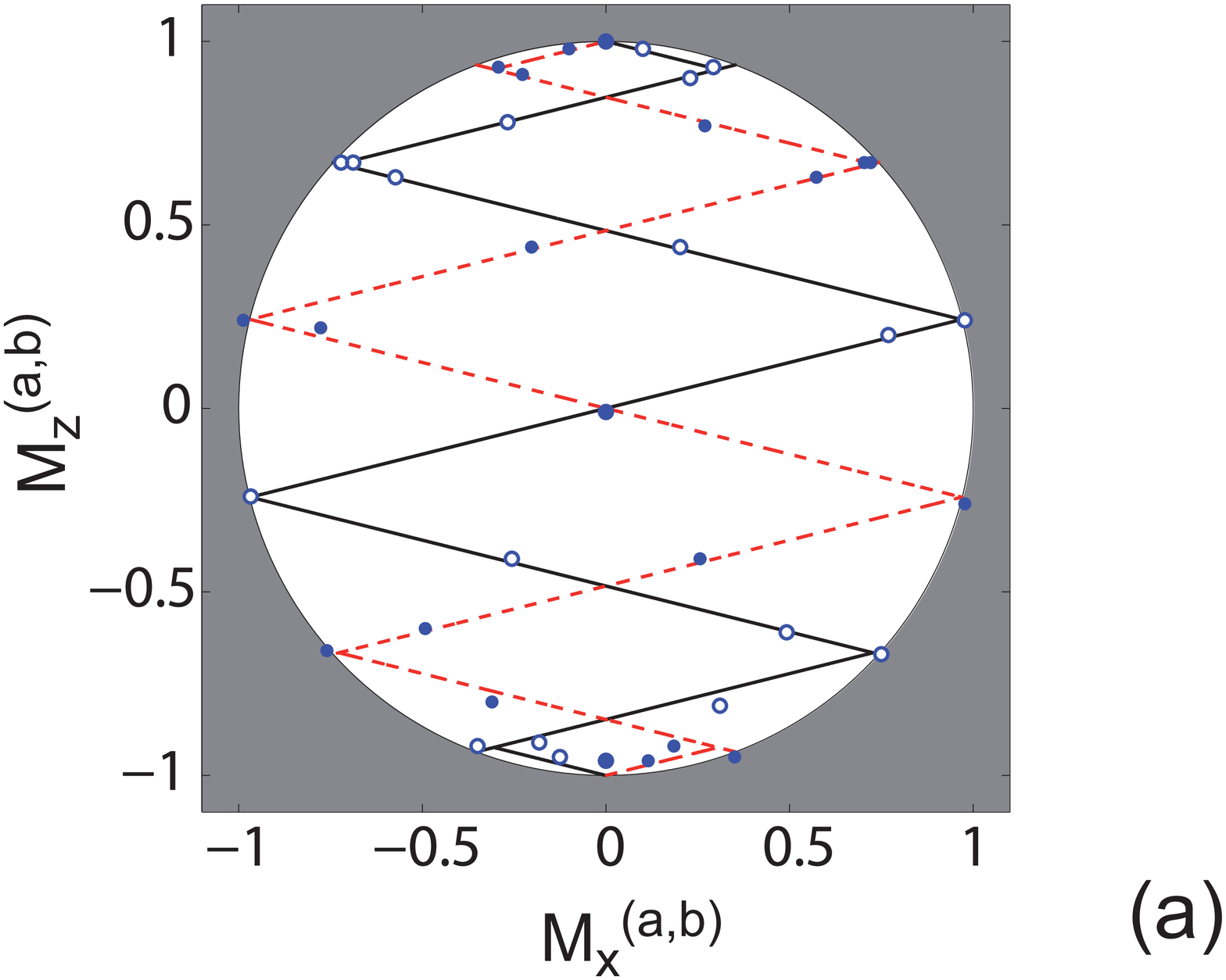}
\includegraphics[height=2in]{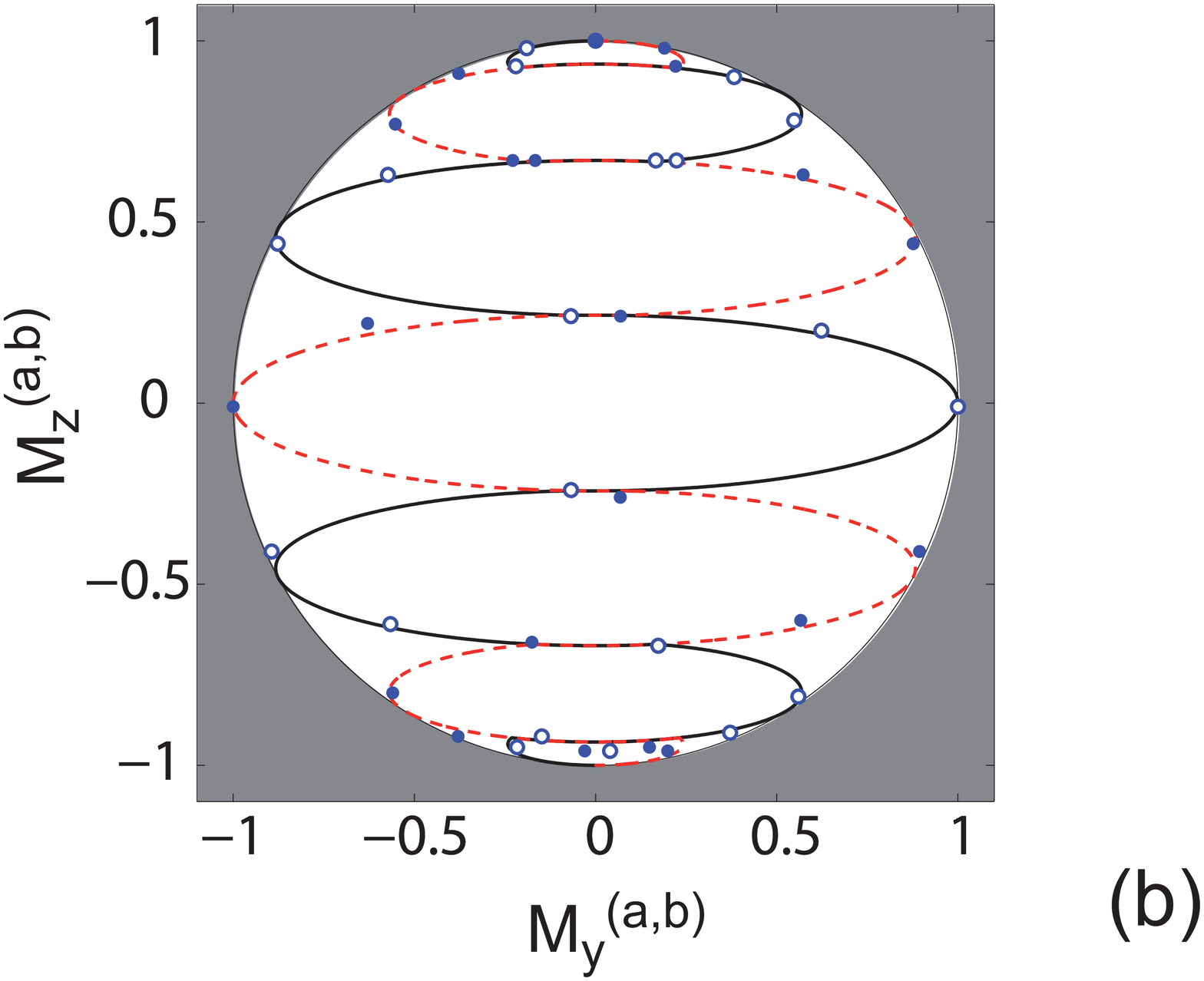}
\end{center}
\caption{(Color Online) Projections of the optimal trajectories for the inversion of the two Bloch vectors with offsets, $\omega/2 \pi= \pm 483$ Hz, in the $(x,z)$- and $(y,z)$- plane are shown in (a) and (b), respectively. The simulated trajectories of the two spins are plotted in black solid line and red dashed line. The experimentally measured trajectories of the two magnetization vectors are represented by open and filled circles.\label{fig2}}
\end{figure}

Here we demonstrate the inversion for the case where the symmetric offsets $\omega_a=-\omega_b$ are four times larger than the maximum radiofrequency (rf) amplitude using the techniques of nuclear magnetic resonance. The optimal pulse (a bang-bang pulse) is shown in Fig. \ref{fig1} and implemented on a Bruker Avance 600MHz spectrometer with linearized amplifiers. The experiment was performed using the two distinct proton spin signals of methyl acetate (dissolved in deuterated chloroform). The two resonances, one from the  $-$OCH$_3$ moiety and the other from the $-$OOCCH$_3$ moiety, were separated by 966 Hz in the $^1$H NMR spectrum. The irradiation frequency was positioned in the center of the two peaks, i.e. $\omega_0=(\omega_a+\omega_b)/2$, resulting in offsets of $\omega=(\omega_a-\omega_b)/2=2 \pi \times 483$ Hz for the two resonances. The maximum rf amplitude was  chosen to be $\omega_{max}=\omega/4=2 \pi \times 120.75$ Hz and the duration of the optimal inversion pulse shown in Fig. \ref{fig1} is $T_p$=6.409 ms. At room temperature (298K), the experimentally measured relaxation time constants of the two spins are $T_1^a \approx T_1^b = 4.95$ s, $T_2^a \approx T_2^b = 140$ ms, which have a negligible effect during the much shorter pulse duration $T_p$. The $x$ and $y$ components of the Bloch vectors, $M_x^{a,b}(t)$ and $M_y^{a,b}(t)$ were measured experimentally by interrupting the optimal pulse shape after the time $t$ and measuring the amplitude and phase of the signal after Fourier transformation of the resulting free induction decay (FID). In order to measure the $z$ component of the Bloch vectors, the experiments were repeated with the addition of a pulsed magnetic field gradient (of duration about 0.2 ms with sine shape), followed by a $90^{\circ}$ hard pulse.
A reasonable match between simulated and experimentally determined trajectories is found. For example,
Fig. \ref{fig1} shows the simulated and experimental trajectories of $M_x^{a}(t)$, $M_y^{a}(t)$ and $M_z^{a}(t)$ as a function of time. Fig. \ref{fig2} shows the projections of the simulated and experimental trajectories of both Bloch vectors.

\section{Summary}\label{sec5}
In this last section, we give a brief overview of the results
obtained in this paper. The four relevant cases for the
simultaneous inversion of two spins are the following.
\begin{enumerate}
\item two control fields along the $x$- and $y$- directions and
one offset $\omega$: the optimal solution is a $\pi$- pulse
\cite{boscain1}.
\item one control field and one offset $\omega$: the optimal
solution is a bang-bang pulse sequence with a number of switching
depending upon the ratio $\omega/\omega_{max}$ \cite{boscain2}.
\item one control field and two offsets $\omega$ and
$-\omega$: The optimal solution is the same as in (2).
\item two control fields and two offsets $\omega$ and
$-\omega$: The optimal solution is also the same as in (2).
\end{enumerate}
\textbf{Aknowledgment.}\\
S.J.G. acknowledges support from
the DFG (GI 203/6-1), SFB 631. S. J. G. and M. B.  thank the Fonds der Chemischen Industrie. Experiments
were performed at the Bavarian NMR center at TU M\"unchen and we thank Dr. R. Marx for helpful discussions and for suggesting the test sample.


\begin{thebibliography}{1}

\bibitem{spin} M. H. Levitt, \emph{Spin dynamics: basics of nuclear magnetic resonance} (John Wiley and sons, New York-London-Sydney, 2008); R. R. Ernst, G. Bodenhausen and A. Wokaun, \emph{Principles of Nuclear Magnetic Resonance in one and two dimensions} (International Series of Monographs on Chemistry, Oxford University Press, Oxford, 1990)

\bibitem{vander}  L. M. K. Vandersypen and I. L. Chuang, Rev. Mod. Phys. \textbf{76}, 1037 (2005);
N. C. Nielsen, C. Kehlet, S. J. Glaser and N. Khaneja, in Encyclopedia of Magnetic Resonance, Ed. R. K. Harris and R. Wasylishen, John Wiley, Chichester (2010).

\bibitem{fruech} D. P. Frueh, T. Ito, J.-S. Li, G. Wagner, S. J.
Glaser and N. Khaneja, J. Bio. NMR. \textbf{32}, 23 (2005); J. L. Neves, B. Heitmann, N. Khaneja and S. J. Glaser, J. Magn. Reson. \textbf{201}, 7 (2009).

\bibitem{chuang}

D. G. Cory, A. F. Fahmy, and T. F. Havel, Proc. Natl. Acad. Sci.
USA  \textbf{94}, 1634(1997);
N.A. Gershenfeld and I.L. Chuang, Science  \textbf{275}, 350 (1997).
%I. Chuang, L. Vandersypen, X. Zhou, D. Leung and S. Lloyd, Nature (London) \textbf{393}, 344 (1998).

\bibitem{nielsen} M. A. Nielsen and I. L. Chuang, \emph{Quantum Computation and Quantum
Information} (Cambridge University Press, Cambridge, 2000).


\bibitem{jurdjevic} V. Jurdjevic, \emph{Geometric control theory} (Cambridge University Press, Cambridge, 1996).

\bibitem{bonnard} B. Bonnard and M. Chyba, \emph{Singular trajectories and their role in control theory} (Springer SMAI, Vol. 40, 2003).

\bibitem{boscain1} U. Boscain, G. Charlot, J.-P. Gauthier, S. Gu\'erin and H. R. Jauslin, J. Math. Phys. \textbf{43}, 2107 (2002)

\bibitem{boscain2} U. Boscain and P. Mason, J. Math. Phys. \textbf{47}, 062101 (2006).

\bibitem{boscainbook} U. Boscain and B. Piccoli, \emph{Optimal syntheses for control systems on 2-D manifolds}, Math\'ematiques and Applications, 43, Springer-Verlag, Berlin, 2004.

\bibitem{bryson} A. E. Bryson and Y.-C. Ho, \emph{Applied Optimal Control}, Taylon and Francis group, New-York-London, 1975.

\bibitem{khaneja} N. Khaneja, R. Brockett and S. J. Glaser, Phys.
Rev. A \textbf{63}, 032308 (2001); N. Khaneja, S. J. Glaser and R.
Brockett, Phys. Rev. A \textbf{65}, 032301 (2002); H. Yuan and N.
Khaneja, Phys. Rev. A \textbf{72}, 040301(R) (2005)

\bibitem{lapertglaser} M. Lapert, Y. Zhang, M. Braun, S. J. Glaser and D. Sugny, Phys. Rev. Lett. \textbf{104}, 083001 (2010).

\bibitem{sugny1} D. Sugny, C. Kontz and H.R. Jauslin, Phys. Rev. A
\textbf{76}, 023419 (2007); B. Bonnard and D. Sugny, SIAM
J. on Control and Optimization, \textbf{48}, 1289 (2009); B.
Bonnard, M. Chyba and D. Sugny, IEEE Transactions on Automatic
control, \textbf{54}, 11, 2598 (2009)

\bibitem{grapeino} T. E. Skinner, T. O. Reiss, B. Luy, N. Khaneja and S. J. Glaser, J. Magn. Reson. \textbf{163}, 8 (2003); T. E. Skinner, T. O. Reiss, B. Luy, N. Khaneja and S. J. Glaser, J. Magn. Reson. \textbf{167}, 68 (2004); K. Kobzar, T. E. Skinner, N. Khaneja, S. J. Glaser and B. Luy, J. Magn. Reson. \textbf{170}, 236 (2004); T. E. Skinner, T. O. Reiss, B. Luy, N. Khaneja and S. J. Glaser, J. Magn. Reson. \textbf{172}, 17  (2005); T. E. Skinner, K. Kobzar, B. Luy, R. Bendall, W. Bermel, N. Khaneja and S. J. Glaser, J. Magn. Reson. \textbf{179}, 241 (2006); N. I. Gershenzon, K. Kobzar, B. Luy, S. J. Glaser and T. E. Skinner, J. Magn. Reson. \textbf{188}, 330 (2007); K. Kobzar, T. E. Skinner, N. Khaneja, S. J. Glaser and B. Luy, J. Magn. Reson. \textbf{194}, 58 (2008).

\bibitem{turinici} G. Turinici and H. Rabitz, Phys. Rev. A \textbf{70}, 063412 (2004); H. Rabitz and G. Turinici, Phys. Rev. A \textbf{75}, 043409 (2007); K. Sundermann, H. Rabitz and R. de Vivie-Riedle, Phys. Rev. A \textbf{62}, 013409 (2000).

\bibitem{mono} W. Zhu, J. Botina and H. Rabitz, J. Chem. Phys.
\textbf{108}, 1953 (1998); W. Zhu and H. Rabitz, J. Chem. Phys.
\textbf{110}, 7142 (1999); D. Sugny, C. Kontz, M. Ndong, Y. Justum, G. Dive
and M. Desouter-Lecomte, Phys. Rev. A \textbf{74}, 043419 (2006);
D. Sugny, M. Ndong, D. Lauvergnat, Y. Justum and M.
Desouter-Lecomte, J. Photochem. Photobiol. A \textbf{190}, 359
(2007); M. Ndong, L. Bomble, D. Sugny, Y. Justum and M. Desouter-Lecomte, Phys. Rev. A \textbf{76}, 043424 (2007).

\bibitem{chitour} Y. Chitour, F. Jean and E. Tr\'elat, SIAM J.
Control Optim. \textbf{47}, 1078 (2008).

\end{thebibliography}
\end{document}